\documentclass[letterpaper, 10 pt, conference]{ieeeconf}  

\IEEEoverridecommandlockouts                              

\usepackage{xcolor}
\usepackage{graphics} 
\usepackage{epsfig} 
\usepackage{mathptmx} 
\usepackage{times} 
\usepackage{amsmath} 
\usepackage{amssymb}  
\usepackage{subcaption}
\usepackage[switch]{lineno}
\usepackage{multirow}
\usepackage{booktabs}
\usepackage{acro}

\DeclareAcronym{DOF}{
short=DOF,
long=degree of freedom
}

\DeclareAcronym{NIR}{
short=NIR,
long=near-infrared
}

\DeclareAcronym{CAD}{
short=CAD,
long=solidworks
}

\DeclareAcronym{GT}{
short=GT,
long=ground truth
}

\DeclareAcronym{BCE}{
short=BCE,
long=binary cross-entropy
}

\DeclareAcronym{MSE}{
short=MSE,
long=mean square error
}

\DeclareAcronym{DSC}{
short=DSC,
long=dice similarity coefficient
}

\DeclareAcronym{DC}{
short=DC,
long=direct current
}

\title{\LARGE \bf VeniBot: Towards Autonomous Venipuncture with Semi-supervised Vein Segmentation from Ultrasound Images}

\author{Yu Chen$^{1}$, Yuxuan Wang$^{1}$, Bolin Lai$^{2}$, Zijie Chen$^{1}$, Xu Cao$^{1}$, Nanyang Ye$^{3}$, Zhongyuan Ren$^{4}$, Junbo Zhao$^{5}$, \\ Xiao-Yun Zhou$^{6}$, Peng Qi$^{1*}$
\thanks{This work is supported by the National Natural Science Foundation of China (Number 51905379) and Shanghai Science and Technology Development Funds (Number 20QC1400900)}%
\thanks{$^{1}$ Yu Chen, Yuxuan Wang, Zijie Chen, Xu Cao and Peng Qi are with Tongji University, Shanghai, China. {\tt\small pqi@tongji.edu.cn}}%
\thanks{$^{2}$ Bolin Lai is with PingAn Technology Co. Ltd., Shanghai, China.}%
\thanks{$^{3}$ Nanyang Ye is with Shanghai Jiao Tong University, Shanghai, China.}%
\thanks{$^{4}$ Zhongyuan Ren is with Soochow University Medical College, Suzhou, China.}%
\thanks{$^{5}$ Junbo Zhao is with Zhejiang University, Hangzhou, China.}%
\thanks{$^{6}$ Xiao-Yun Zhou is with PAII Inc., MD, USA.}%
\thanks{$^{*}$ corresponding author.}%
}

\begin{document}

\maketitle
\thispagestyle{empty}
\pagestyle{empty}

\begin{abstract}

In the modern medical care, venipuncture is an indispensable procedure for both diagnosis and treatment. In this paper, unlike existing solutions that fully or partially rely on professional assistance, we propose VeniBot --- a compact robotic system solution integrating both novel hardware and software developments. For the hardware, we design a set of units to facilitate the supporting, positioning, puncturing and imaging functionalities. For the software, to move towards a full automation, we propose a novel deep learning framework --- semi-ResNeXt-Unet for semi-supervised vein segmentation from ultrasound images. From which, the depth information of vein is calculated and used to enable automated navigation for the puncturing unit. VeniBot is validated on 40 volunteers, where ultrasound images can be collected successfully. For the vein segmentation validation, the proposed semi-ResNeXt-Unet improves the \ac{DSC} by $5.36\%$, decreases the centroid error by 1.38 pixels and decreases the failure rate by $5.60\%$, compared to fully-supervised ResNeXt-Unet.

\end{abstract}

\section{INTRODUCTION}
As a popular invasive procedure for venous access, the venipuncture has been widely used and acknowledged for clinical evaluation and treatment.
This procedure opens up broad clinical practices including but not limited to biochemical analysis of the blood sample, infusion of fluid and cannulation for catheterization.
For most adults, venipuncture is conventionally performed by medical practitioners under direct visual monitorization.
Despite its simplicity, the success rate of venipuncture is potentially influenced by a variaty of factors.
In some adverse cases, the rate can be below 50\% under improper settings~\cite{mbamalu1999methods}.
On one hand, veipuncture is an experience-based procedure requiring extensive practices, which is hard for novices. On the other hand, in specific populations such as the elderly or infants, people with dark complexion, or patients in shock state, veins are not easily accessible. In addition, medical practitioners are at a high risk of needle injury and infection of blood-borne diseases like hepatitis and human immunodeficent virus (HIV) \cite{mengistu2021worldwide} or airborne diseases like corona virus disease \cite{scoppettuolo2020vascular}. Hence, novel solutions are necessitated to provide better and safe healthcare service.

Robotics has been a popular paradigm to automate many traditional manual tasks, for example, autonomous driving \cite{levinson2011towards,grigorescu2020survey,kiran2020deep} and autonomous surgical operation \cite{thai2020advanced,ma2020autonomous,zhou2020application}. Some robots have been developed and widely used for veinpuncture. For example, Veebot, which was founded in 2010, proposed a venipuncture robot that combines a robotic arm and a compact puncturing unit \cite{perry2013profile}. It identifies a coarse vein with the help of infrared light and examines whether the vein is suitable or not for puncture with ultrasound. Venouspro was proposed by Vasculogic. It contains a six \ac{DOF} positioning unit and a three \ac{DOF} distal manipulator \cite{chen2020deep}. It is more compact and portable, compared to the robotic arm proposed by Veebot. MagicNurse, which is an automatic blood drawing robot, draws the blood and exchanges the puncture needle automatically. It detects the forearm vein with NIR images and estimates the depth of the needle with force sensors. The device has more than 20 \ac{DOF} and is as big as a refrigerator. KUKA AG 
invented a robot arm holding a wireless ultrasound probe as an automatically robot-assisted tracking system to scan the vessel. It uses active contours to segment the vessel and it works on phantoms \cite{unger2020robot}.

\begin{figure}[tbp]
    \centering
    \includegraphics[width=0.5\textwidth]{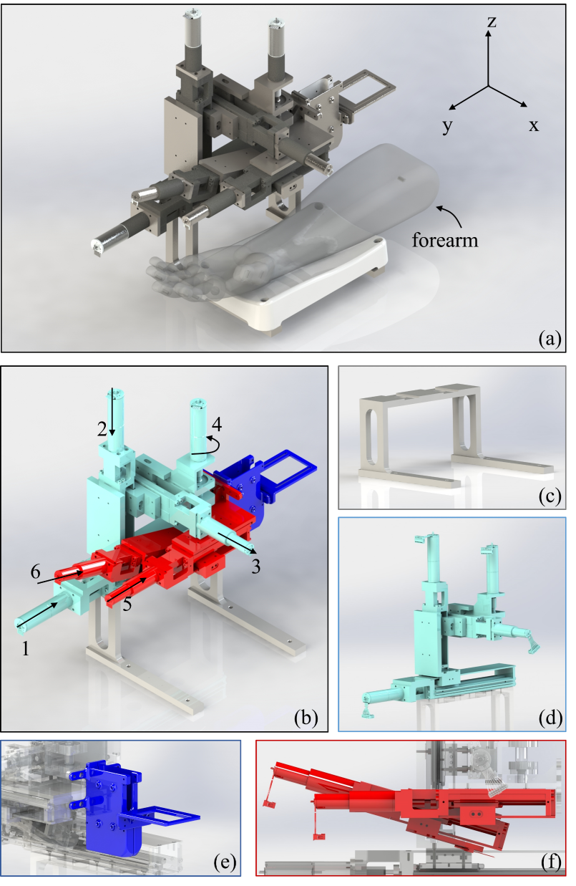}
    \caption{An illustration of the \ac{CAD} model of the proposed VeniBot, with the white color indicating the supporting unit, the blue color indicating the positioning unit, the red color indicating the puncturing unit, and the purple color indicating the imaging unit.}
    \label{fig:venibot}
\end{figure}

In this paper, a compact venipuncture device - VeniBot is designed and built for autonomous venipucture. The \ac{CAD} model of VeniBot is shown in Fig.~\ref{fig:venibot}. It is consisted of four units, including the supporting unit used to support the whole robot, the positioning unit mainly used to move VeniBot to the puncture target, the puncturing unit mainly used to puncture, and the imaging unit used to mount the \ac{NIR} camera and ultrasound device. The positioning unit contains three single axis robots driven with \ac{DC} motors and one motor that drives directly, and is navigated by a \ac{NIR} camera mounted on the imaging unit, while the puncturing unit contains two single axis robots, and is navigated by an ultrasound device mounted on the imaging unit.

Except the hardware development, software for localization and navigation is also important and essential for robotic automation, typical examples are in medical robotics \cite{zhang2020robotic,zhou2018real} and general robotics \cite{pachon2020robotic}. In the developed VeniBot, there are two navigation problems: (1) the puncture area determination from the \ac{NIR} images; (2) the vein segmentation from the ultrasound images. In this paper, we mainly focus on the second challenge while the first challenge is focused by another IROS submission. Deep learning has been a popular method for image segmentation, however, it relies heavily on the training data with carefully manual labels. A more efficient way is to fully use the unlabeled data and train with semi-supervised learning. Popular semi-supervised methods include $\rm{\Pi}$-model \cite{laine2016temporal}, temporal ensemble \cite{laine2016temporal}, mix match \cite{berthelot2019mixmatch} and mean teacher \cite{tarvainen2017mean}. In this paper, inspired by mean teacher \cite{tarvainen2017mean} and Unet \cite{ronneberger2015u}, we propose semi-ResNeXt-Unet for vein segmentation from limited labeled training data and large amount of unlabeled training data.

The two main novelties and contributions of this paper are:

\begin{itemize}
    \item VeniBot is proposed and built with four units (supporting unit, positioning unit, puncturing unit and imaging unit) and six motors (four motors for the positioning unit and two motors for the puncturing unit).
    \item Semi-ResNeXt-Unet is proposed to segment the vein from ultrasound images with semi-supervised learning, from which, the vein center is extracted to navigate the VeniBot to puncture automatically.
\end{itemize}

We validate the proposed VeniBot on puncture paper with $2mm \times 2mm$ grids and the proposed semi-ResNeXt-Unet on 40 volunteers. The later achieves a \ac{DSC} of $75.11\%$, a centroid error of 8.23 pixels and a failure rate of $2.00\%$ on 10 labeled training volunteers and 30 unlabeled training volunteers, which indicates a promising performance of the proposed methods.

In the following context, we will articulate the hardware of VeniBot in details in Sec.~\ref{sec:hardware}. Afterwards, we'll detailedly describe the automatic puncturing unit of VeniBot in Sec.~\ref{sec:puncture}, the validation of VeniBot and puncturing unit is in Sec. \ref{sec:result}, and the conclusion is in Sec. \ref{sec:conclusion}.

\section{METHODOLOGY}
\label{sec:method}

\subsection{Hardware Design of VeniBot}
\label{sec:hardware}

The first unit of VeniBot is the supporting unit which has zero \ac{DOF}, as indicated in white color in Fig.~\ref{fig:venibot}. The main function of supporting unit is to support and hold the whole VeniBot. On top of the supporting unit, we assemble the positioning unit which is with four \ac{DOF}s. The main function of positioning unit is to predict the most suitable venipuncture position and angle in the xOy plane, under the navigation of a NIR camera.


\begin{figure}[tbp]
    \centering
    \includegraphics[width=0.5\textwidth]{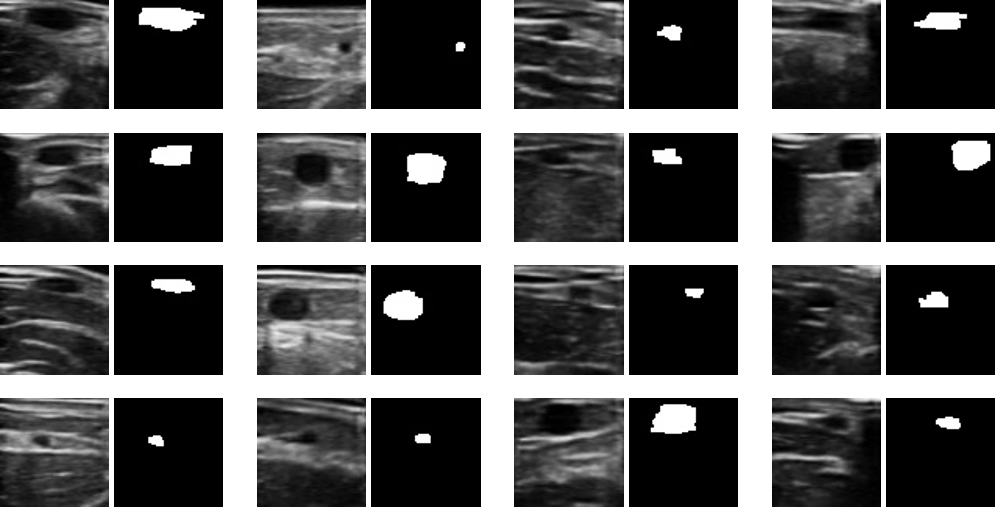}
    \caption{An illustration of the ultrasound images (left) and the corresponding \ac{GT} masks (right).}
    \label{fig:us}
\end{figure}

On top of the positioning unit, we assemble the third unit of VeniBot -- the puncturing unit which has two DoFs, as indicated in red color in Fig.~\ref{fig:venibot}. The main function of the puncturing unit is to puncture the vein under the navigation of an ultrasound scanner. The puncturing unit is consisted of two single axis robots, where both of them have one DoF of translation. A ultrasound device (ST-1C transducer with frequency of 7.5MHz, 48 lateral array elements, 80 axial array elements, element spacing of 0.3mm) is mounted at the
imaging unit. At the beginning, the positioning unit moves to the best venipuncture position. Then motor 2 moves downward until a high-quality cross-section view of the vein is obtained from the ultrasound scanner which persistently scans the skin during the entire moving period. The segmentation mask and center position of the vein are localized automatically using the deep learning method --- semi-ResNeXt-Unet proposed in Sec.~\ref{sec:puncture}, which is trained with limited labeled data and a large amount of unlabeled data. The puncturing unit contains two \ac{DOF}s: motor 5 drives a y-direction single axis robot and motor 6 drives another single axis robot, which is at an angle of about $17^\circ$ from the xOy plane. Motor 6 is fixed on motor 5 and moves along the y direction when motor 5 works. The needle is fixed on single axis robot 6 and moves along the direction of it when motor 6 works. The depth of the vein’s centroid guides motor 5 to move single axis robot 6. This step will finally determine the needle depth. Then motor 6 works and pushes the needle tip to the vein centroid. In this paper, we mainly focus on the automation of puncturing unit. While the automation of positioning part is illustrated in the other IROS submission.

\subsection{Automatic Vein Segmentation and Center Position Determination}
\label{sec:puncture}

\begin{figure}[tbp]
    \centering
    \includegraphics[width=0.5\textwidth]{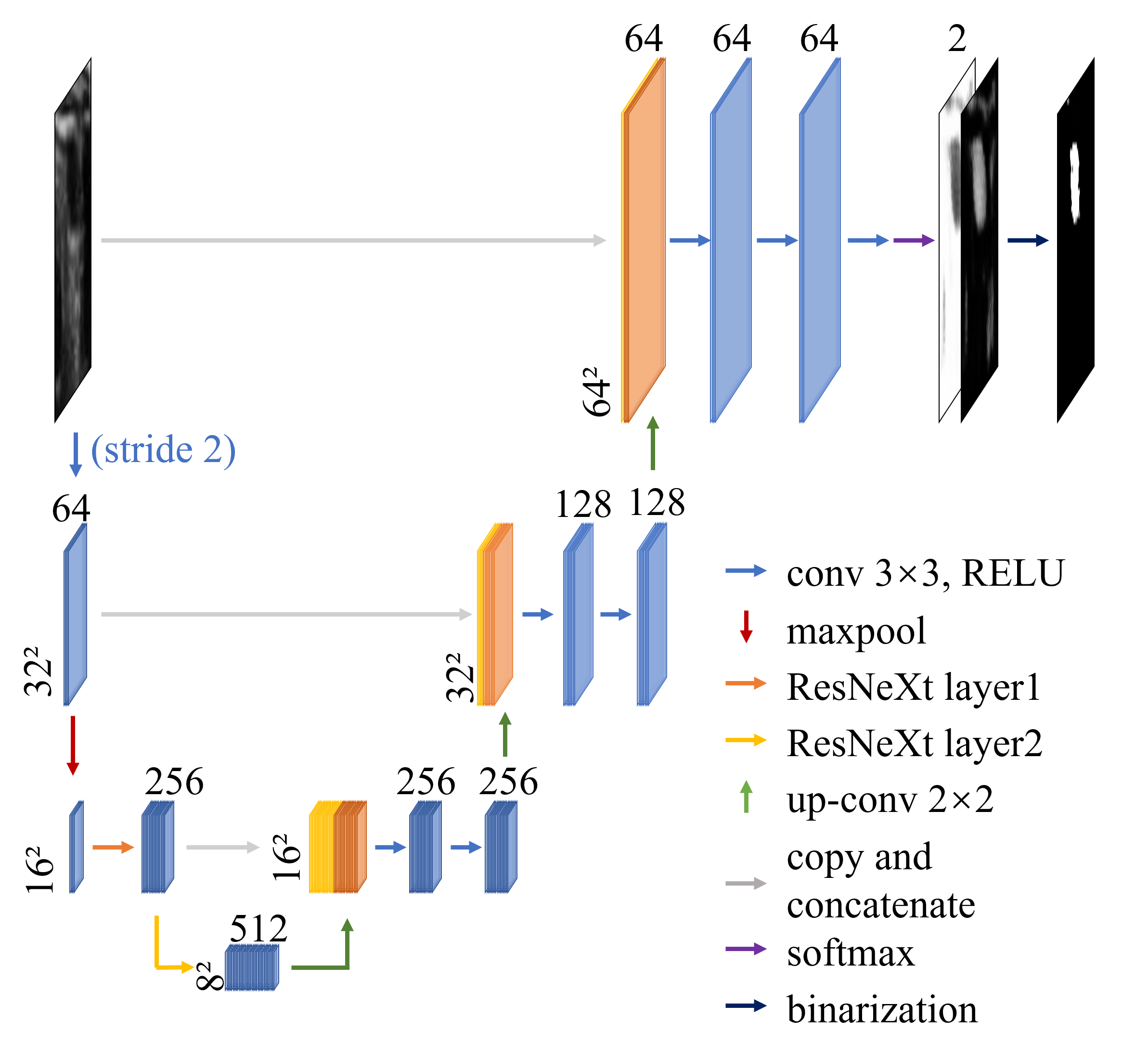}
    \caption{An illustration of the proposed ResNeXt-Unet network structure used for vein segmentation.}
    \label{fig:network}
\end{figure}

To achieve autonomous venipuncture with the proposed VeniBot, it's of great importance to develop an automatic method to localize centers of veins from ultrasound images, as it indicates the depth of the movement of motor 5 and 6. In this paper, centers of veins are calculated from their segmentation masks\footnote{Following~\cite{zhou2018towards}, segmentation masks are used to localize centers of targets rather than detection bounding boxes because pixel-wise masks contain more information of shape and hence can further supply more accurate center location.}. However, vein segmentation from ultrasound images is very challenging, as the image quality is susceptible to the experience of operators and low quality usually exists even scanned by well-trained operators. Another challenge is the deformation caused by transducer. Since ultrasound wave can not transit through the air, the transducer has to touch the patient’s forearm skin with suitable pressure. Large pressure makes the vein squashed, while small pressure remains some air between the skin and the transducer, resulting in all-black ultrasound images. Even for well-obtained ultrasound images, the vein bound can be obscure and its shape can differ dramastically, indicating the difficulty of labeling large amounts of ultrasound images manually. See Fig.~\ref{fig:us} for some examples.

To address the time-consuming and labor-intensive annotations, we propose a semi-supervised segmentation method --- semi-ResNeXt-Unet which first trains a fully-supervised model on limited labeled data and then improves its performance on more unlabeled data. Semi-ResNeXt-Unet contains 10 convolutional layers and 4 deconvolutional layers as demonstrated in Fig.~\ref{fig:network} and Table~\ref{tab:network}.

\begin{table}[tbp]
\caption{The layer details of the proposed ResNeXt-Unet network structure used for vein segmentation.}
\centering
\begin{tabular}{c|c|c}
\hline
stage&output&ResNeXt-Unet\\ 
\hline
conv1&$32\times32$&$3\times3$, 64, stride 2\\
\hline
\multirow{2}{*}{conv2}
&\multirow{2}{*}{$16\times16$}
&$3\times3$ maxpool, stride 2\\
\cline{3-3}
&&$\left[\begin{array}{l}
     1\times1, 128  \\
     3\times3, 128, C=32\\
     1\times1, 256
\end{array}\right]\times3$\\
\hline
conv3&$8\times8$
&$\left[\begin{array}{l}
     1\times1, 256  \\
     3\times3, 256, C=32\\
     1\times1, 512
\end{array}\right]\times3$\\
\hline
deconv4&$16\times16$&$3\times3$, 256, stride 2\\
\hline
\multirow{2}{*}{conv5}
&\multirow{2}{*}{$16\times16$}
&concatenate\\
\cline{3-3}
&&$3\times3$, 256, stride 1\\
\hline
conv6&$16\times16$&$3\times3$, 256, stride 1\\
\hline
deconv7&$32\times32$&$3\times3$, 128, stride 2\\
\hline
\multirow{2}{*}{conv8}
&\multirow{2}{*}{$32\times32$}
&concatenate\\
\cline{3-3}
&&$3\times3$, 128, stride 1\\
\hline
conv9&$32\times32$&$3\times3$, 128, stride 1\\
\hline
deconv10&$64\times64$&$3\times3$, 64, stride 2\\
\hline
\multirow{2}{*}{conv11}
&\multirow{2}{*}{$64\times64$}
&concatenate\\
\cline{3-3}
&&$3\times3$, 64, stride 1\\
\hline
conv12&$64\times64$&$3\times3$, 64, stride 1\\
\hline
conv13&$64\times64$&$3\times3$, 2, stride 1\\
\hline
\end{tabular}
\label{tab:network}
\end{table}

The semi-supervised method is based on the state-of-the-art mean teacher framework~\cite{tarvainen2017mean}. Assume $M$ labeled images and $N$ unlabeled images are available. Thus we have $\mathbf{S}=\{x_i, y_i\}_{i=1}^M$ and $\mathbf{U}=\{x_i\}_{i=1}^N$ for labeled data and unlabeled data, respectively. $x_i$ and $y_i$ denote input images and pixel-wise annotations. As illustrated in Fig.~\ref{fig:meanteacher}, the proposed semi-ResNeXt-Unet works in the following steps:

\begin{figure}
    \centering
    \includegraphics[width=0.5\textwidth]{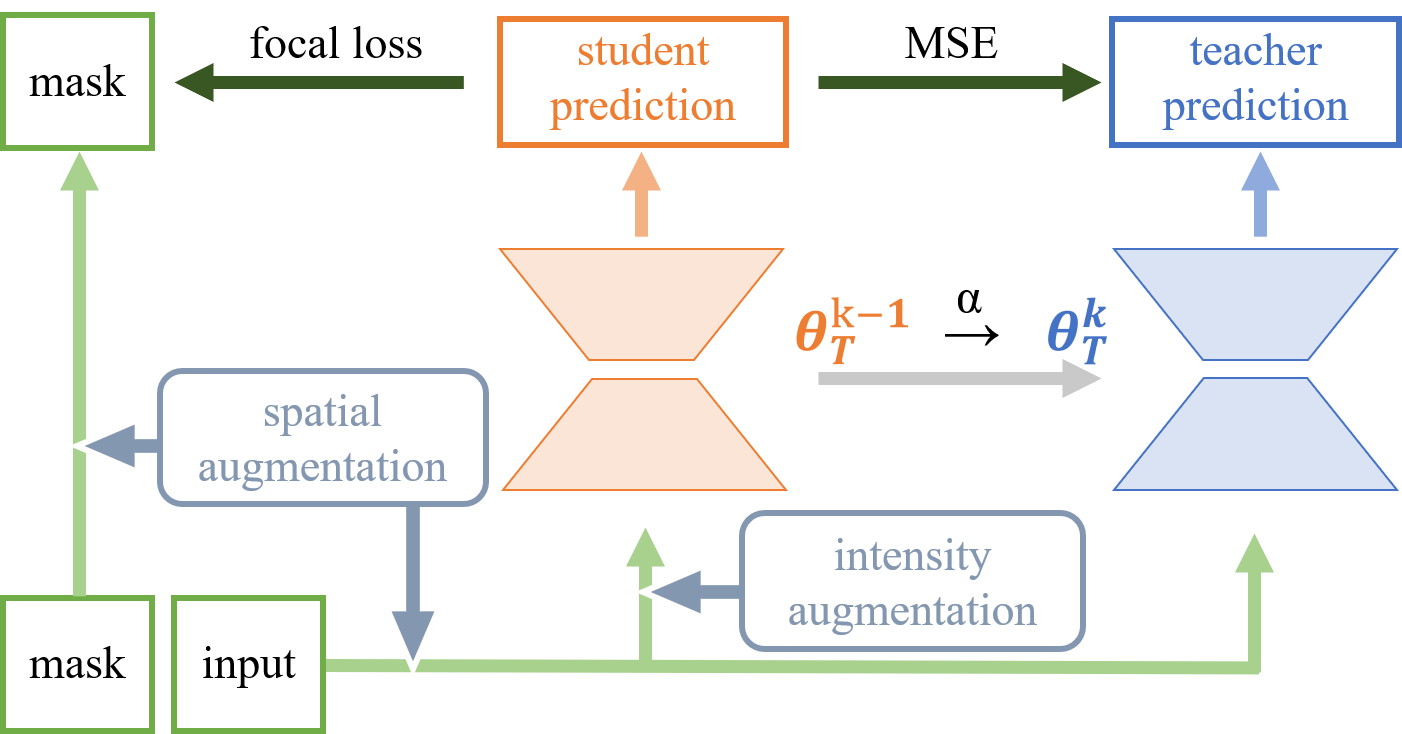}
    \caption{An illustration of the semi-ResNeXt-Unet framework for semi-supervised learning.}
    \label{fig:meanteacher}
\end{figure}

(1) Train a fully-supervised network based on $\mathbf{S}$. The \ac{BCE} loss is used at first. When reasonable prediction masks are obtained from the network. Focal loss \cite{lin2017focal} is used in place of 
\ac{BCE} loss to address the extreme imbalance of foreground and background in ultrasound images. After the convergence, the parameters are used as initialization for the teacher and student models.

(2) Input labeled images $\mathbf{S}$ into the student model. Focal loss is still used to train the network with other training protocols the same as that used for the fully-supervised training.

(3) For unlabeled images $\mathbf{U}$, raw images are augmented by spatial transformation and input into the teacher model to get reliable pseudo probability maps. Then they are further disturbed by extra intensity augmentation and input into the student model. We use \ac{MSE} as the consistency loss to penalize any difference between teacher and student model's outputs. Then the total loss can be expressed as a weighted summation of supervised and semi-supervised losses.
\begin{equation}
    L_{total}=\frac{\lambda_1 M \mathcal{L}_{sup} + \lambda_2 N \mathcal{L}_{semisup}}{M+N},
\end{equation}
where the $\lambda_1$ and $\lambda_2$ are hyper-parameters used to balance the two losses.

(4) In the $k$ iteration, parameters of the teacher model ($\theta_T$) are frozen at first and the student model's parameters ($\theta_S$) are updated by back-propagation of the focal loss and MSE loss. Then teacher model's parameters are updated by moving average from the student model according to the following equation
\begin{equation}
    \theta_T^k = \alpha \theta_T^{k-1} + (1-\alpha) \theta_S^k,
\end{equation}
where $\alpha$ is a hyper-parameter to control the pace of update. After training, either the teacher model or the student model, which shows better performance on the validation set, is used for inference. 

After segmenting the area of vein, to minimize the center detection error of the network and make sure the target vein is safe to access, a group of operation and evaluation is necessary. The image after binarization will go trough opening operation to remove outliers and closing operation to link broken regions of the vein. We remain the biggest connected component and filter out other foreground areas. The target for venipuncture is the centroid of the segmented vein mask, which can be calculated by averaging the x coordinates and y coordinates of all pixels in the segmentation mask, respectively. It can be expressed as 
\begin{equation}
    (x_c,y_c)=(\frac{\sum_{i=1}^n x_i}{n},\frac{\sum_{i=1}^n y_i}{n}),
\end{equation}
where $(x_c,y_c)$ and $(x_i,y_i)$ are coordinates of the centroid of the vein and $i$-th pixel in the mask. $n$ is the total number of pixels. $(x_c, y_c)$ further determines the final position of axis 5 and axis 6.

\begin{figure}
    \centering
    \includegraphics[width=0.5\textwidth]{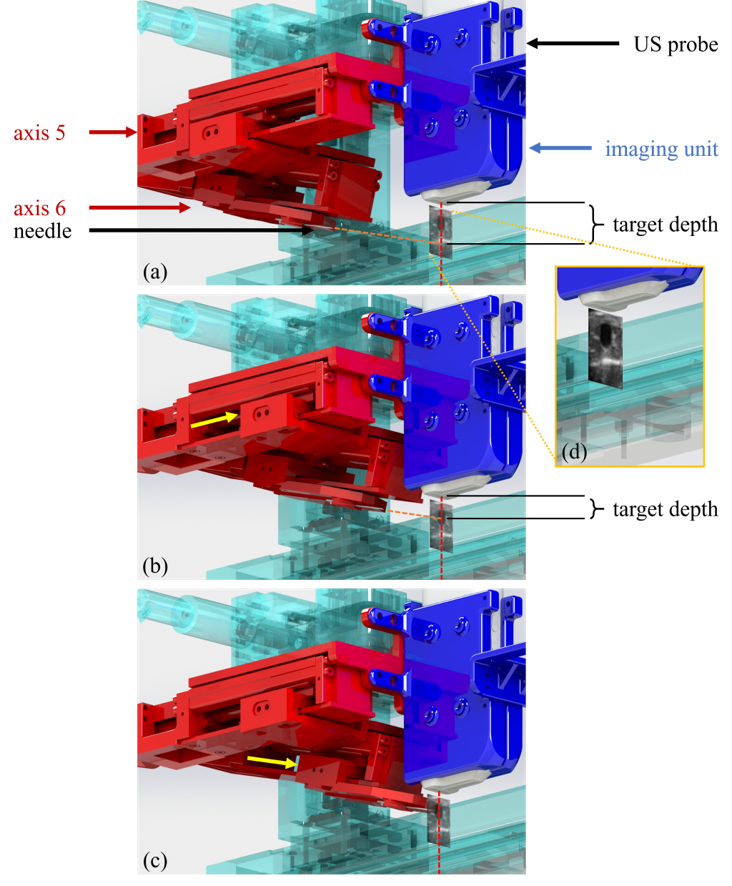}
    \caption{An illustration of the puncturing unit works under the guidance of centroid coordinates. (a) The ultrasound scanner first scans the forearm and gets the image of longitudinal section, which includes the vein. (b) Then motor 5 changes the final puncture target depth of the needle that is mounted on the end of axis 6. (c) After the predicted centroid is aligned with the real one of vein, motor 6 works and sends the needle into the vein.}
    \label{fig:puncture}
\end{figure}

\section{VALIDATION}
\label{sec:result}

\subsection{Experiments}

The real VeniBot and the experimental setup are demonstrated in Fig.~\ref{fig:realveni}. The supporting unit was assembled with aluminium pieces. The skeleton of positioning unit and puncturing unit were built with machine work, while the base of ultrasound probe and NIR camera was processed with 3D metal printing. All of the translation pairs were driven by the combination of DC motors and single axis robots. Axis 4, a rotation pair, was directly driven by the DC motor. In total, four HIWIN KK40 single aixis robots, one HIWIN KK50 single axis robot, three EC-max16 DC motors and three EC-max22 DC motors from Maxon Motor Inc. are used in VeniBot. Note that all ultrasound images used in this paper were collected by our machine, validating the robustness and stability of the proposed VeniBot.

Painting liquid or solid ultrasonic coupling agent on the skin is necessary to drive away the air and keep the ultrasound wave propagating in solid or liquid media. From the aspect of computer graphic method, the difference between the vein and the background and the characteristics of vein section are obvious enough to distinguish the vein from background. The intensity of vein is low because blood, in the state of liquid, does not reflect the ultrasound wave. Moreover, the vein section is an approximate circle or ellipse. Most of the veins are at the size of 1$\sim$2mm in diameter. These characteristics above can be manually measured with intensity, circularity, size, convexity, concavity and aspect ratio, indicating the availability of manual labeling of \ac{GT}. We use VeniBot to collect ultrasound images from 40 volunteers (27 males and 13 females) at the age of 18$\sim$70 to validate our methods. Each of them provided 30 ultrasound images of forearm vein, \emph{i.e.} 1200 images in total, with the same size of $70\times70$. By adjusting the parameters of multiple filters to segment the vein area one by one, we segmented 300 images as labeled training data and used images from the rest 30 volunteers as unlabeled data. 

Two kinds of augmentations are implemented in fully- and semi-supervised training: spatial augmentation and intensity augmentation. Spatial augmentation includes resizing, flipping, rotation, shearing and changing the aspect ratio. Intensity augmentation includes randomly adding or multiplying a certain value, contrast normalization and dropout. The range of the spatial and intensity augmentation is listed in Table~\ref{tab:aug}.

\begin{table}[tbp]
\caption{The range of the spatial and intensity augmentation added on the training of the semi-ResNeXt-Unet model.}\label{tab:aug}
\centering
\begin{tabular}{c|c|c}
\hline
\multicolumn{2}{c|}{augmentation}&size, range or probability \\
\hline
\multirow{5}{*}{spatial } & resizing & to $64\times64$ \\
\cline{2-3}
                          & horizontal flipping & 50\% probability \\
\cline{2-3}
                          & rotation & $-15^{\circ}\sim15^{\circ}$ \\
\cline{2-3}
                          & shearing & $-15^{\circ}\sim15^{\circ}$ \\
\cline{2-3}
                          & changing aspect ratio & $-0.01\sim0.01$ \\
\hline
\multirow{4}{*}{intensity }& adding  & -15$\sim$15 \\
\cline{2-3}
                           & multiplying & $0.8\sim1.2$\\
\cline{2-3}
                          & dropout & $0\sim0.05$ \\
\cline{2-3}
                          & contrast normalization & $0.8\sim1.2$ \\
\hline
\end{tabular}
\end{table}

The ResNeXt-Unet model is trained with 300 labeled images first. The model is first trained using \ac{BCE} loss and with the learning rate of $1\times10^{-3}$. Then it is further trained using focal loss with a learning rate of $5\times10^{-4}$ until convergence. The batch size is set as 8 and both intensity and spatial augmentations are used. In the semi-supervised training, labeled images and unlabeled images are randomly selected in each batch. The same spatial augmentations are used on student and teacher model's inputs, while intensity augmentations are only implemented on student model's inputs.


\begin{table*}[!htbp]
\centering
\caption{The mean and std DSC of training the fully-supervised model, pseudo label training model, $\rm{\Pi}$-model, temporal ensemble model and the proposed semi-ResNeXt-Unet model on the five cross validations.}
\begin{tabular}{cc|c|c|c|c|c|c}
\hline
\multicolumn{2}{c|}{\multirow{2}{*}{experiment}}&\multicolumn{6}{c}{DSC}\\
\cline{3-8}
&&model 1&model 2&model 3&model 4&model 5&average\\
\hline
\multicolumn{2}{c|}{fully-supervised}&0.6917$\pm$0.3565&0.6507$\pm$0.3760&0.7339$\pm$0.2949&0.7435$\pm$0.2817&0.7088$\pm$0.3135&0.6975$\pm$0.3245\\
\hline

\multicolumn{2}{c|}{pseudo label training \cite{sohn2020simple}}&0.6749$\pm$0.3697&0.6294$\pm$0.3930&0.7128$\pm$0.3197&0.7148$\pm$0.3115&0.6757$\pm$0.3446&0.6724$\pm$0.3477\\
\hline
\multicolumn{2}{c|}{$\rm{\Pi}$-model \cite{laine2016temporal}}&0.7205$\pm$0.3150&0.6634$\pm$0.3636&0.7608$\pm$0.2677&0.7814$\pm$0.2547&0.7337$\pm$0.2831&0.7320$\pm$0.2972\\
\hline
\multicolumn{2}{c|}{temporal ensemble \cite{laine2016temporal}}&0.6689$\pm$0.3576&0.6136$\pm$0.4016&0.7057$\pm$0.3235&0.7281$\pm$0.3047&0.6480$\pm$0.3630&0.6618$\pm$0.3501\\
\hline
\multirow{2}{*}{semi-supervised}&
\textbf{student}&
\textbf{0.7291$\pm$0.3092}&
\textbf{0.6792$\pm$ 0.3441}&
\textbf{0.7749$\pm$0.2517}&
\textbf{0.8138$\pm$0.2149}&
\textbf{0.7584$\pm$0.2584}&
\textbf{0.7511$\pm$0.2557}\\
\cline{2-8}
&teacher&0.7041$\pm$0.3245&0.6730$\pm$0.3478&0.7599$\pm$0.2682&0.7718$\pm$0.2623&0.7446$\pm$0.2706&0.7307$\pm$0.2947\\
\hline
\end{tabular}
\label{tab:dscerror}
\end{table*}

\begin{table*}[!htbp]
\centering
\caption{The mean and std centroid error of training the fully-supervised model, pseudo label training model, $\rm{\Pi}$-model, temporal ensemble model and the proposed semi-ResNeXt-Unet model on the five cross validations.}
\begin{tabular}{cc|c|c|c|c|c|c|c}
\hline
\multicolumn{2}{c|}{\multirow{2}{*}{experiment}}&\multicolumn{6}{c|}{centroid MSE error}&\multirow{2}{*}{vein radius}\\
\cline{3-8}
&&model 1&model 2&model 3&model 4&model 5&average\\
\hline
\multicolumn{2}{c|}{fully-supervised}&14.19$\pm$18.48&\textbf{12.40$\pm$14.80}&8.39$\pm$12.50&5.57$\pm$8.26&7.50$\pm$9.96&9.61$\pm$12.80&\multirow{6}{*}{14.22$\pm$4.38}\\
\cline{1-8}

\multicolumn{2}{c|}{pseudo label training \cite{sohn2020simple}}&14.81$\pm$16.77&15.52$\pm$16.56&9.21$\pm$12.87&6.60$\pm$9.06&8.56$\pm$11.54&10.94$\pm$13.36\\
\cline{1-8}
\multicolumn{2}{c|}{$\rm{\Pi}$-model \cite{laine2016temporal}}&\textbf{7.26$\pm$13.40}&14.40$\pm$14.01&7.27$\pm$10.56&5.21$\pm$7.79&\textbf{7.47$\pm$9.53}&8.32$\pm$11.06\\
\cline{1-8}
\multicolumn{2}{c|}{temporal ensemble \cite{laine2016temporal}}&10.99$\pm$15.30&14.93$\pm$16.16&7.90$\pm$10.08&5.85$\pm$7.96&12.75$\pm$12.51&10.48$\pm$12.40\\
\cline{1-8}
\multirow{2}{*}{semi-supervised}&\textbf{student}&8.78$\pm$14.09&13.20$\pm$ 12.77&\textbf{7.07$\pm$10.58}&\textbf{4.30$\pm$7.65}&7.80$\pm$10.34&\textbf{8.23$\pm$11.09}\\
\cline{2-8}
&teacher&8.08$\pm$12.28&13.05$\pm$12.12&7.80$\pm$10.50&4.86$\pm$7.66&9.88$\pm$10.78&8.73$\pm$10.67\\
\hline
\end{tabular}
\label{tab:mseerror}
\end{table*}

\begin{table*}[!htbp]
\centering
\setlength{\tabcolsep}{5mm}
\caption{The mean failure rate of training the fully-supervised model, pseudo label training model, $\rm{\Pi}$-model, temporal ensemble model and the proposed semi-ResNeXt-Unet model on the five cross validations.}
\begin{tabular}{cc|c|c|c|c|c|c}
\hline
\multicolumn{2}{c|}{\multirow{2}{*}{experiment}}&\multicolumn{6}{c}{failure rate}\\
\cline{3-8}
&&model 1&model 2&model 3&model 4&model 5&average\\
\hline
\multicolumn{2}{c|}{fully-supervised}&21.67\%&6.67\%&5.00\%&3.00\%&1.67\%&7.60\%\\
\hline

\multicolumn{2}{c|}{pseudo label training \cite{sohn2020simple}}&25\%&21.67\%&8.33\%&3.33\%&8.33\%&13.33\%\\
\hline
\multicolumn{2}{c|}{$\rm{\Pi}$-model \cite{laine2016temporal}}&10.00\%&6.67\%&6.67\%&0.00\%&0.00\%&4.67\%\\
\hline
\multicolumn{2}{c|}{temporal ensemble \cite{laine2016temporal}}&\textbf{1.67\%}&11.67\%&5.00\%&1.67\%&1.67\%&4.34\%\\
\hline
\multirow{2}{*}{semi-supervised}&
student&6.67\%&3.33\%&0.00\%&0.00\%&0.00\%&2.00\%\\
\cline{2-8}
&\textbf{teacher}&
5.00\%&
\textbf{0.00\%}&
\textbf{0.00\%}&
\textbf{0.00\%}&
\textbf{0.00\%}&
\textbf{1.00\%}\\
\hline
\end{tabular}
\label{tab:failureerror}
\end{table*}

\subsection{Results}

\begin{figure}
    \centering
    \includegraphics[width=0.5\textwidth]{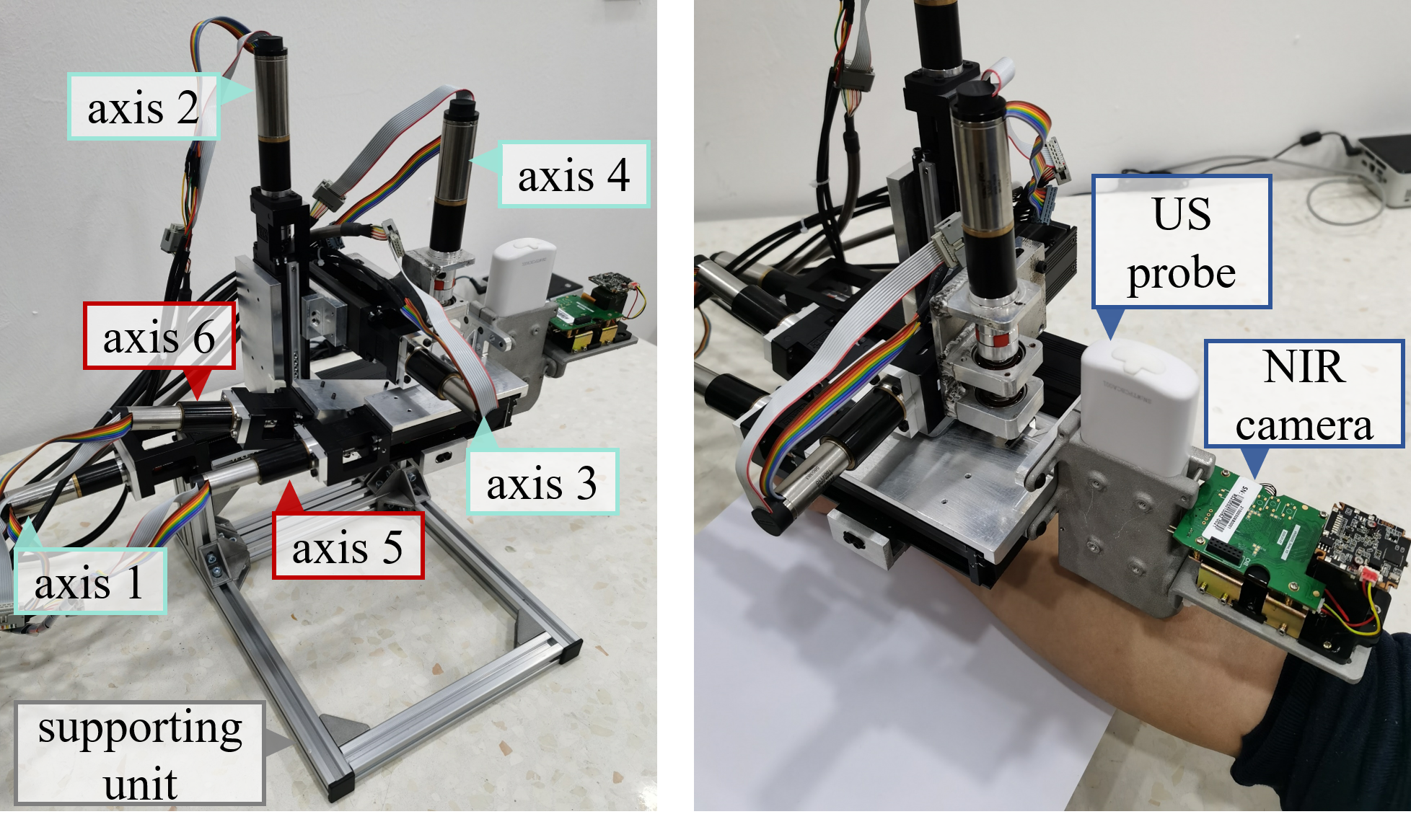}
    \caption{An illustration of the real VeniBot and the process of ultrasound image collection.}
    \label{fig:realveni}
\end{figure}

\begin{figure}
    \centering
    \includegraphics[width=0.5\textwidth]{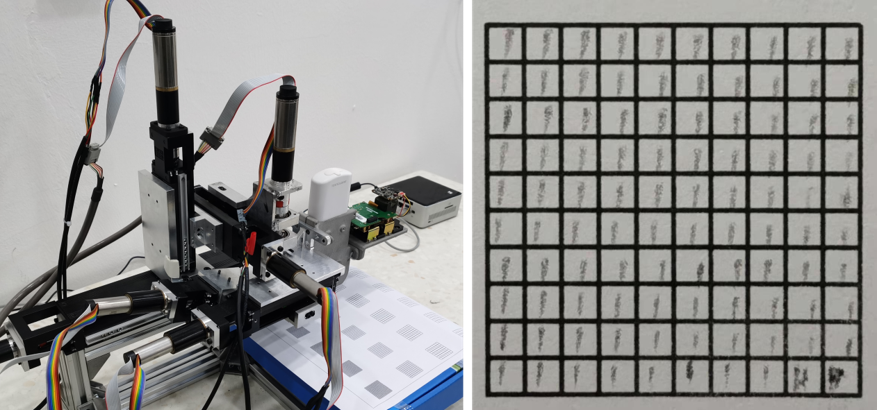}
    \caption{An illustration of the kinematic and assembly accuracy validation experiment.}
    \label{fig:realveni}
\end{figure}

\begin{figure}
    \centering
    \includegraphics[width=0.5
    \textwidth]{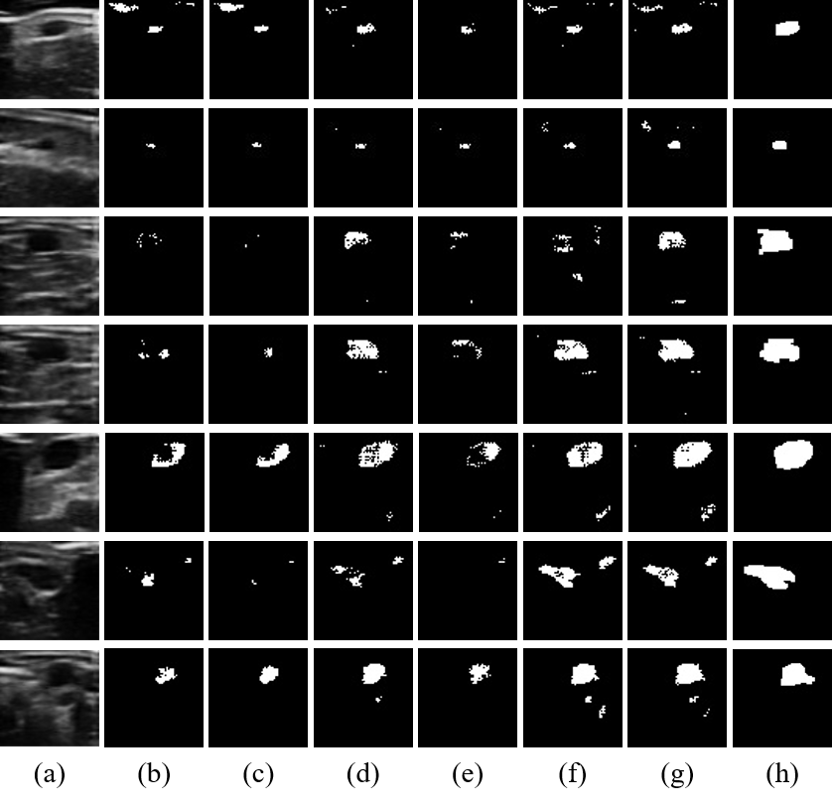}
    \caption{Illustration of (a) input image, and segmentation results of (b) fully-supervised model, (c) pseudo label training, (d) $\rm{\Pi}$-model; (e) temporal ensemble, (f) teacher and (g) student model in semi-ResNeXt-Unet.}
    \label{fig:result}
\end{figure}

To validate the kinematic and assembly accuracy of Venibot, a 10$\times$10 grid paper with a grid size of 2mm$\times$2mm is used. We assembled a pencil at the end of axis 6 and controlled Venibot to move 2mm along axis x or y once at a time. After each movement, the pencil moved along axis 6 to draw a short line in the corresponding grid. As shown in Fig.~\ref{fig:realveni}, VeniBot successfully controlled a pencil drawing short lines inside each grid, which shows the crucial and fundamental ability of VeniBot to achieve safe and viable venipuncture.

In order to show the advantage of the proposed semi-ResNeXt-Unet, three strong semi-supervised baselines are used: (1) pseudo label training \cite{sohn2020simple} which generates pseudo \ac{GT} for the unlabeled data with the converged model in the fully-supervised step, (2) $\rm{\Pi}$-model \cite{laine2016temporal} which only uses one model with the consistency loss, temporal ensemble \cite{laine2016temporal} which updates the pseudo GT after every training epoch using moving average method. The DSC of our method and baselines with 5-fold cross validation is illustrated in Tab.~\ref{tab:dscerror}. It can be observed that temporal ensemble and pseudo label training achieve the lowest performance, which is even lower than the fully-supervised model. One possible reason is that the challenging nature of vein segmentation causes lots of error in the pseudo \ac{GT} from the fully-supervised model, which misleads the model in semi-supervised training. In contrast, based on the consistency loss, $\rm{\Pi}$-model and our method boost the performance by a significant margin. The proposed semi-ResNeXt-Unet still shows its superiority and outperforms the $\rm{\Pi}-model$ by $1.91\%$. Compared with the fully-supervised counter-part, the mean \ac{DSC} is finally improved by $5.36\%$.

We further show the centroid \ac{MSE} error which measures the distance between the centroid of prediction and \ac{GT}. If there is no vein predicted, we count this situation as failure cases and see its prediction at the top left corner. The centroid \ac{MSE} error and the failure rate are shown in Tab. \ref{tab:mseerror} and \ref{tab:failureerror} respectively. For the centroid error, the proposed semi-ResNeXt-Unet outperforms the fully-supervised method, pseudo label training, and temporal ensemble with notable margins, while outperforms $\pi$-model with a small margin. For the failure rate, the proposed semi-ResNeXt-Unet achieves the lowest value for four of the five cross validations and the overall average failure rate, indicating the robustness of the proposed semi-ResNeXt-Unet to hard cases, \textit{i.e.}, images with small veins.

Finally, seven vein segmentation results, segmented by the four methods, were randomly selected and shown in Fig.~\ref{fig:result}. The results of baselines lose some significant vein areas or even totally fail to generate reasonable masks. The proposed semi-ResNeXt-Unet precisely catches the vein and outputs more reliable masks, which is an important perceptual evaluation of the superiority of our method.

\section{CONCLUSION}
\label{sec:conclusion}

In the paper, we proposed a compact venipuncture robot -- VeniBot. For its automation, we proposed ResNeXt-Unet to automatically segment the vein from ultrasound images and calculate the depth from skin to the vein centroid. Given the expense of collecting a large labeled dataset, we also proposed a semi-supervised segmentation network -- semi-ResNeXt-Unet to obtain reasonable vein masks based on a limited set of labeled training data as well as more unlabeled images. It prominently improves the performance by $5.36\%$ \ac{DSC} compared with the fully-supervised counter-part and surpasses other semi-supervised baselines by at least $1.91\%$ \ac{DSC}. It's a promising step towards a wide and automatic application in the future.

\clearpage

\bibliographystyle{IEEEtran}
\bibliography{IROS.bib}

\end{document}